\documentclass{svproc}
\usepackage{url}

\usepackage{graphicx}% Include figure files
\usepackage{bm}% bold math
\usepackage{hyperref}

\begin{document}
\mainmatter              % start of a contribution
\title{Charged and Neutral Current Pion Production in Neutrino-Nucleus Scattering}

\titlerunning{Coherent Pion Production}

\author{Kapil Saraswat $^{1}$ \and Prashant Shukla $^{2,3}$ \and Vineet Kumar $^{2}$ 
\and Venktesh Singh $^{1}$}

\authorrunning{Kapil Saraswat et al.} % abbreviated author list (for running head)

\institute{$^{1}$ Department of Physics, Banaras Hindu University, Varanasi 221005, India \\ 
$^{2}$ Nuclear Physics Division, Bhabha Atomic Research Centre, Mumbai 400085, India \\
$^{3}$ Homi Bhabha National Institute, Anushakti Nagar, Mumbai 400094, India}

\maketitle              % typeset the title of the contribution

\begin{abstract}
{\footnotesize In this article, we present the charged and neutral current coherent 
pion production in the neutrino-nucleus interaction in the resonance region using 
the formalism based on the partially conserved axial current (PCAC) theorem which 
relates the neutrino-nucleus cross section to the pion-nucleus elastic cross section. 
The pion nucleus elastic cross section is calculated using the Glauber model approach. 
We calculate the integrated cross sections for neutrino-carbon, neutrino-iron and 
neutrino-oxygen scattering. The results of integrated cross-section calculations are 
compared with the measured data.}
\end{abstract}

\section{Introduction}
{\footnotesize The neutrinos in the range between 1 to 3 GeV can interact with matter 
by many processes such as quasi elastic scattering (QES), interaction via resonance 
pion production (RES) and deep inelastic scattering (DIS). One of the most important 
processes in the RES region is coherent pion production. Coherent pion production is 
a process where the nucleus interact as a whole with neutrino and remain in the same 
quantum state as it was initially when the neutrino arrived.}

\section{PCAC Based Model}
{\footnotesize The PCAC based approach is used to calculate the differential cross 
section for the charged and neutral coherent pion production process which s given 
in Ref. \cite{Kopeliovich:1992ym} \cite{Saraswat:2016kln}. The used kinematic cuts 
are given in Ref.\cite{Saraswat:2016kln}. $\xi$ is a parameter which is used for the 
pion dominating picture in the scattering process. The pion-nucleus cross section is 
based on the Glauber model approach which is described in Ref.\cite{Saraswat:2016kln}.}

\section{Results and Discussions}

\begin{figure}[!htb]
\minipage{0.33\textwidth}
\includegraphics[width=\linewidth]{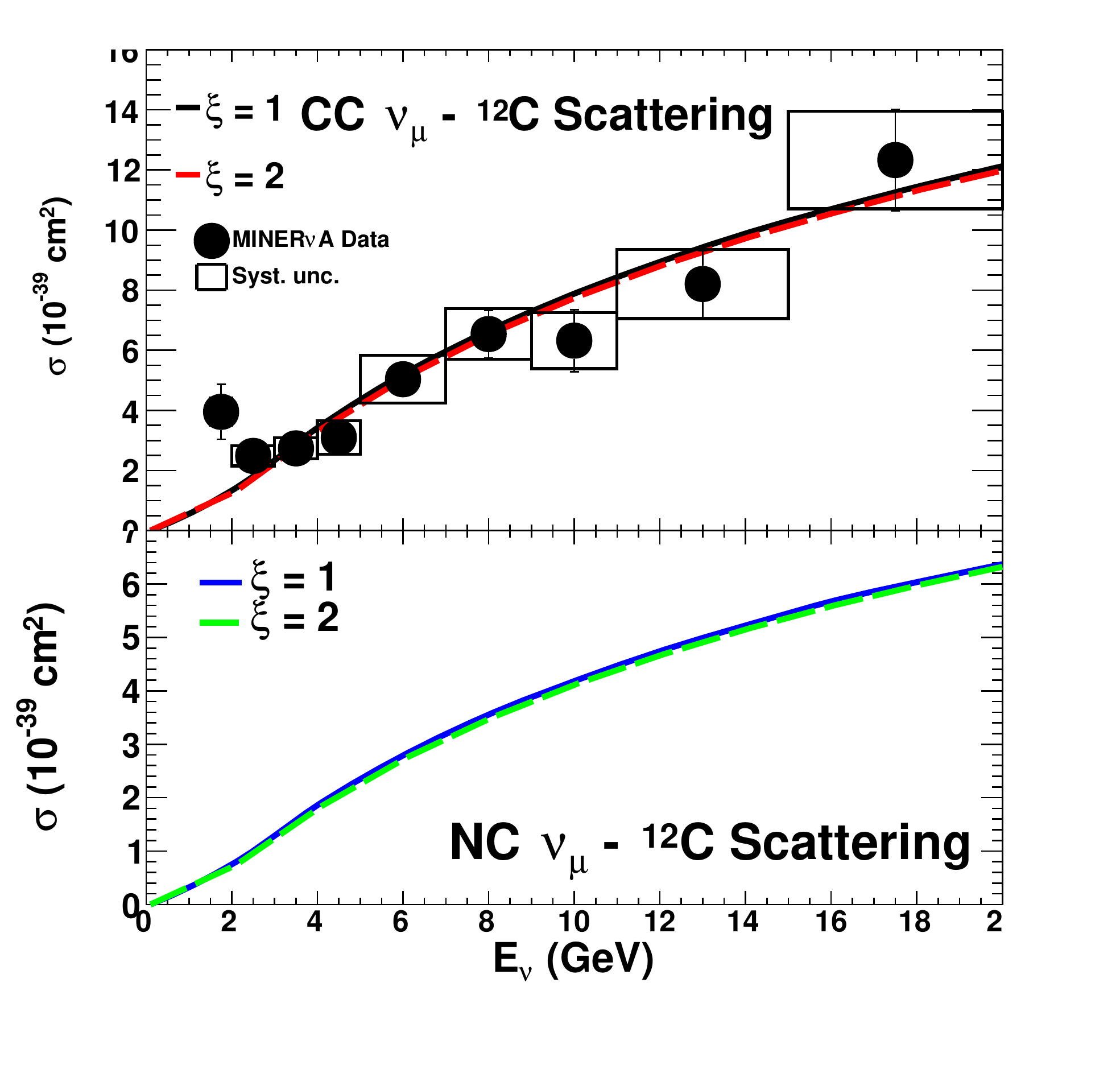}
\caption{{\footnotesize $\sigma$ of $\nu_{\mu}-^{12}C$ scattering as a function 
of $E_{\nu}$.}}
\label{figure1_pion_carbon}
\endminipage
\hfill
\minipage{0.33\textwidth}
\includegraphics[width=\linewidth]{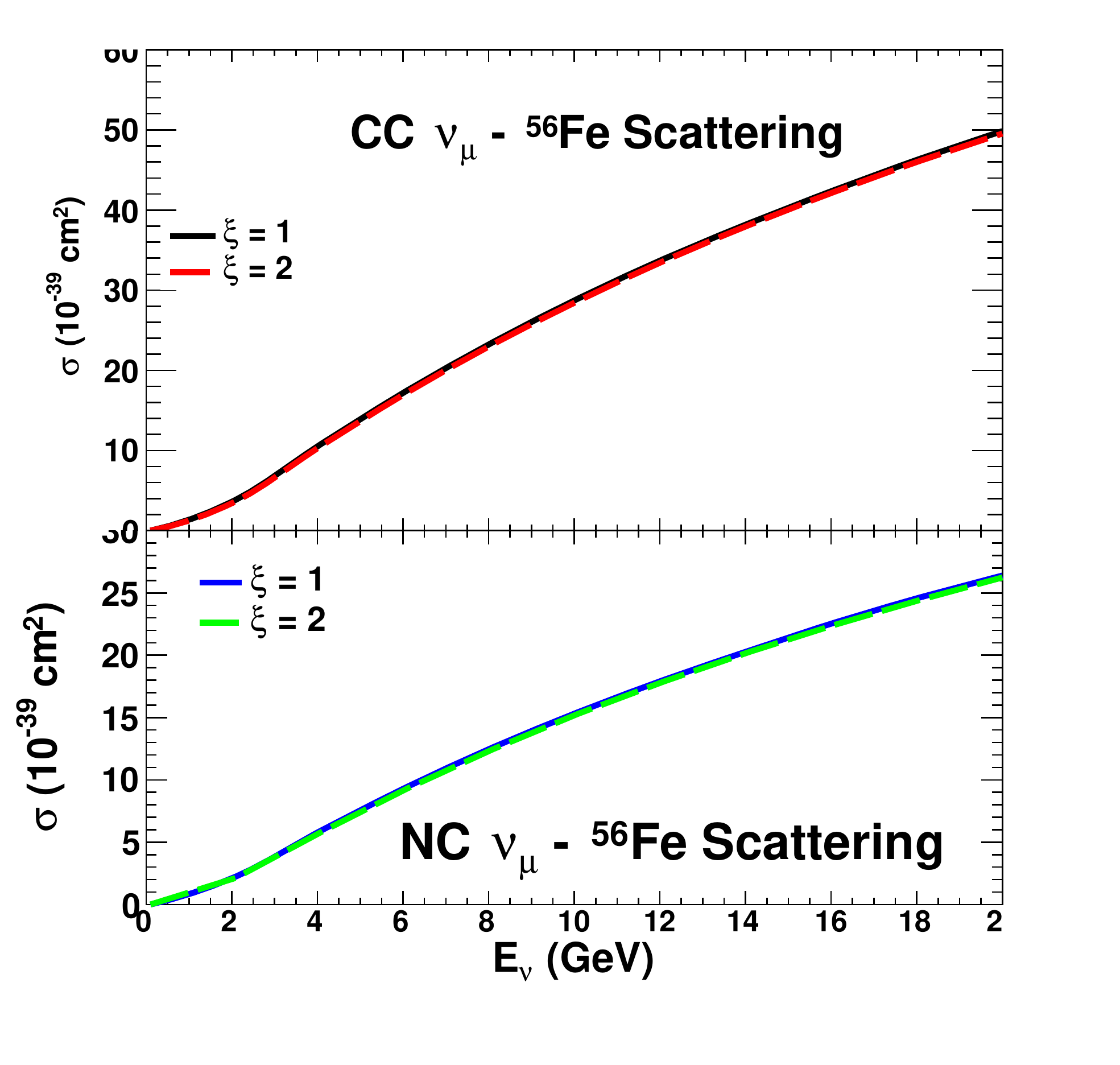}
\caption{{\footnotesize $\sigma$ of $\nu_{\mu}-^{56}Fe$ scattering as a function 
of $E_{\nu}$}.}
\label{figure2_pion_iron}
\endminipage
\hfill
\minipage{0.33\textwidth}%
\includegraphics[width=\linewidth]{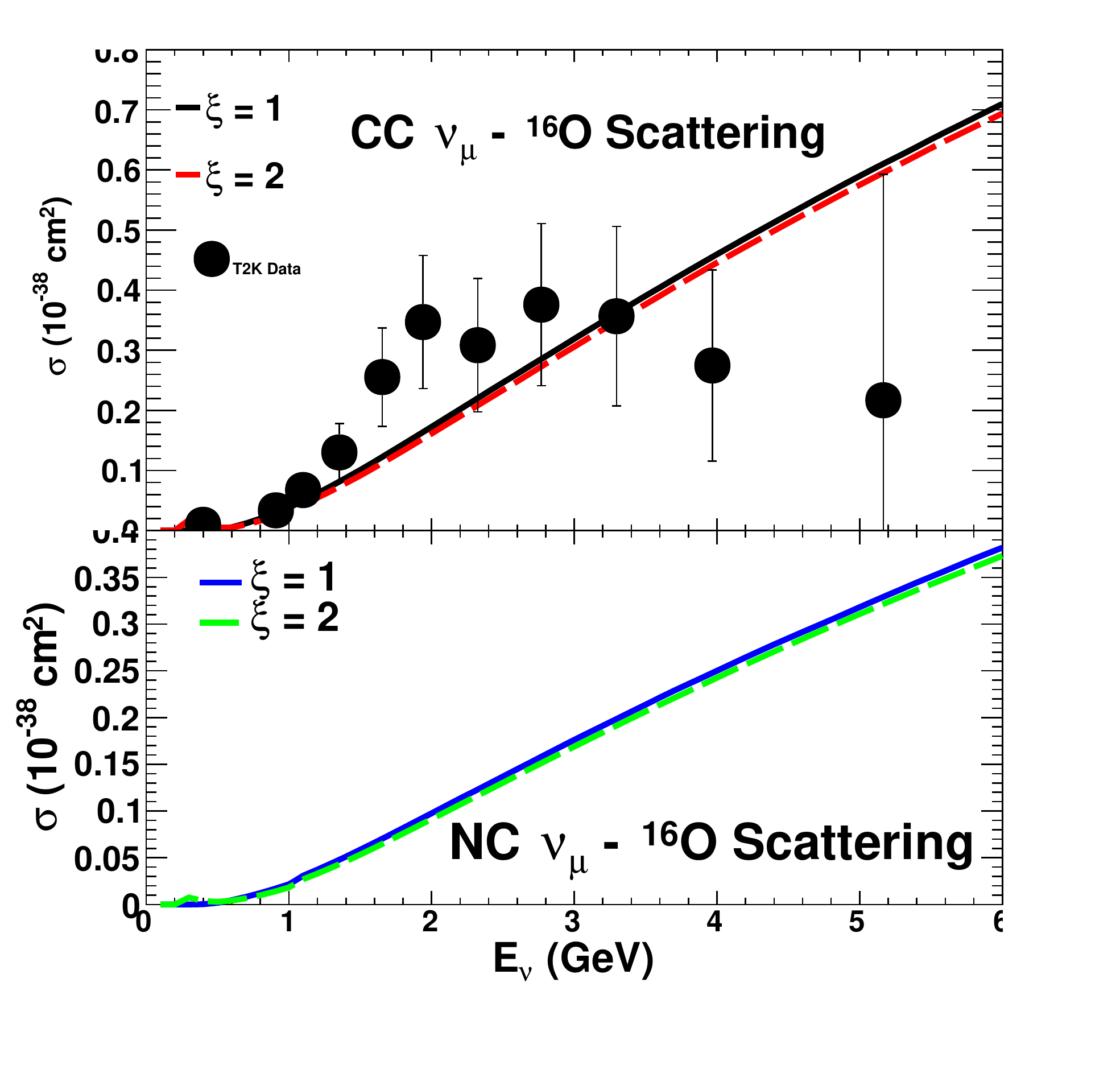}
\caption{{\footnotesize $\sigma$ of $\nu_{\mu}-^{16}O$ scattering as a function 
of $E_{\nu}$.}}
\label{figure3_pion_oxygen}
\endminipage
\end{figure}

{\footnotesize Figure \ref{figure1_pion_carbon} shows the total cross section 
$\sigma$ for the neutrino-carbon scattering as a function of the neutrino energy 
$E_{\nu}$ obtained using the PCAC model for $\xi$ = 1 and 2. The upper panel shows 
the charged current (CC) case while the lower panel shows the neutral current (NC). 
In the upper panel, the calculations are compared with the MINER$\nu$A data 
\cite{Higuera:2014azj}. The calculations are compatible with the data. In both panels, 
the cross section is reduced with increasing the value of $\xi$.}
{\footnotesize Figure \ref{figure2_pion_iron} shows the $\sigma$ for the neutrino-iron 
scattering as a function of $E_{\nu}$ obtained using the PCAC model for $\xi$ = 1 and 2.}
{\footnotesize Figure \ref{figure3_pion_oxygen} shows the $\sigma$ for the 
neutrino-oxygen scattering as a function of $E_{\nu}$ obtained using the PCAC model 
for $\xi$ = 1 and 2. In the upper panel, the calculations are compared with the 
T2K data \cite{Abe:2016aoo}. The calculations are compatible with the data at 
lower neutrino energy. With increasing the value of $\xi$, the total cross section 
is reduced.}

\section{Conclusion}
{\footnotesize We presented the integrated cross section for the charged and neutral 
coherent pion production in neutrino - nucleus scattering using the formalism based 
on PCAC theorem which relates the neutrino - nucleus cross section to the pion - nucleus 
elastic cross section. We study the behavior of the cross section as a function of neutrino 
energy and the parameters of the model. The calculations are compared with the experimental 
data. The calculation gives good description of the MINER$\nu$A data while it gives good 
description of the T2K data at low neutrino energy.}

%
% ---- Bibliography ----
%


\begin{thebibliography}{6}

\bibitem{Kopeliovich:1992ym}
  B.~Z.~Kopeliovich and P.~Marage,
  %``Low Q**2, high neutrino nu physics (CVC, PCAC, hadron dominance),''
  Int.\ J.\ Mod.\ Phys.\ A {\bf 8} (1993) 1513.

\bibitem{Saraswat:2016kln}
  K.~Saraswat, P.~Shukla, V.~Kumar and V.~Singh,
  %``Coherent pion production in neutrino-nucleus scattering,''
  Phys.\ Rev.\ C {\bf 93} (2016) 035504. %[arXiv:1602.07820 [hep-ph]].

\bibitem{Higuera:2014azj}
  A.~Higuera {\it et al.} [MINERvA Collaboration],
  %``Measurement of Coherent Production of $\pi^\pm$ in Neutrino and Antineutrino Beams on Carbon from $E_\nu$ of $1.5$ to $20$ GeV,''
  Phys.\ Rev.\ Lett.\  {\bf 113} (2014) 261802. %[arXiv:1409.3835 [hep-ex]].

\bibitem{Abe:2016aoo}
  K.~Abe {\it et al.} [T2K Collaboration],
  %``First measurement of the muon neutrino charged current single pion production cross section on water with the T2K near detector,''
  Phys.\ Rev.\ D {\bf 95} (2017) 012010. %[arXiv:1605.07964 [hep-ex]].

\end{thebibliography}
\end{document}